\documentclass[final,5p,times,twocolumn]{elsarticle}

\usepackage{amssymb}
\usepackage{lipsum}
\usepackage{graphicx} 
\usepackage{amsmath}
\usepackage{float}    
\usepackage{caption}
\usepackage{enumitem}

\journal{Nuclear Instruments and Methods in Physics Research Section A}

\begin{document}

\begin{frontmatter}



\title{DeepMuon: Accelerating Cosmic-Ray Muon Simulation Based on Optimal Transport}


\author[1]{\normalsize Ao-Bo Wang}
\author[1]{Chu-Cheng Pan}
\author[1]{Xiang Dong}
\author[1]{Yu-Chang Sun}
\author[1]{Yu-Xuan Hu}
\author[1]{Ao-Yan Cheng}
\author[1]{Hao Cai\corref{cor1}} 
\ead{hcai@whu.edu.cn} 
\author[1]{Xi-Long Fan\corref{cor2}} 
\ead{xilong.fan@whu.edu.cn}

\affiliation[1]{organization={School of Physics and Technology, Wuhan University},
            city={Wuhan},
            postcode={430072},
            country={People's Republic of China}}

\cortext[cor1]{Corresponding author}
\cortext[cor2]{Co-corresponding author}

\begin{abstract}
    Cosmic muon imaging technology is increasingly being applied in various fields. However, simulating cosmic muons typically requires the rapid generation of a large number of muons and tracking their complex trajectories through intricate structures. This process is highly computationally demanding and consumes significant CPU time.To address these challenges, we introduce DeepMuon, an innovative deep learning model designed to efficiently and accurately generate cosmic muon distributions. In our approach, we employ the inverse Box-Cox transformation to reduce the kurtosis of the muon energy distribution, making it more statistically manageable for the model to learn. Additionally, we utilize the Sliced Wasserstein Distance (SWD) as a loss function to ensure precise simulation of the high-dimensional distributions of cosmic muons.We also demonstrate that DeepMuon can accurately learn muon distribution patterns from a limited set of data, enabling it to simulate real-world cosmic muon distributions as captured by detectors. Compared to traditional tools like CRY, DeepMuon significantly increases the speed of muon generation at sea level. Furthermore, we have developed a pipeline using DeepMuon that directly simulates muon distributions in underwater environments, dramatically accelerating simulations for underwater muon radiography and tomography.For more details on our open-source project, please visit https://github.com/wangab0/deepmuon.
\end{abstract}




\end{frontmatter}




\section{Introduction}
\label{introduction}

Cosmic rays consist mainly of atomic nuclei, with approximately 86\% being hydrogen nuclei (protons), 12\% helium nuclei, and around 1\% carbon, nitrogen, oxygen, and iron nuclei. Additionally, about 1\% of cosmic rays are electrons, with smaller amounts of high-energy gamma rays, neutrinos, and antiprotons. When these primary cosmic rays collide with atoms in the atmosphere, they produce a large number of secondary particles, primarily pions and kaons. Charged pions rapidly decay into muons and neutrinos, while charged kaons decay into pions and subsequently muons. Neutral kaons can decay directly into muons. At sea level, approximately 80\% of the charged particles in cosmic rays are muons, which hit the Earth’s surface at a frequency of about 170 Hz per square meter\cite{dorman2013cosmic}.

As detector technologies have advanced, cosmic-ray muons have found a wide range of applications. For example, L.W. Alvarez used cosmic-ray muons to probe the interior of the pyramids, confirming the absence of unknown chambers of significant size\cite{alvarez1970search}. This method, known as muon radiography or muon absorption radiography, infers the material composition of an object by measuring the muon flux passing through it. It is commonly used for imaging large structures. Another technique, muon tomography, was developed by a team at Los Alamos in 2003\cite{borozdin2003radiographic}. This method uses the scattering angle of muons to create a three-dimensional image of an object by comparing the incoming and outgoing muon beams. 
Both muon radiography and muon tomography rely heavily on simulations of cosmic ray muon trajectories as they pass through different materials. Currently, there are two main types of simulation tools used for cosmic ray muon generation:
\begin{enumerate}
    \item Monte Carlo-based tools, such as CORSIKA\cite{heck2012corsika} and CRY\cite{CRY}, simulate cosmic ray muon production by modeling the physical processes involved. These simulations include the generation of primary cosmic rays, the decay of secondary cosmic particles, and their interactions with the atmosphere. While this method is flexible and allows adjustments to the physical models and parameters, it is computationally expensive due to the complexity of simulating the entire process. The output includes not only muons but other cosmic ray particles as well.
    \item Parameter-based tools, such as CMSCGEN\cite{CMSCGEN} and EcoMug\cite{EcoMug}, rely on empirical or semi-empirical formulas derived from experimental data to describe muon flux under specific conditions. These models can either directly use these formulas or fit Monte Carlo data with polynomials, as in the case of CMSCGEN, which fits data generated by CORSIKA. These parameter-based methods offer faster simulations by focusing specifically on muons and forgoing the simulation of the full range of cosmic ray particles.
\end{enumerate}

Muon trajectory simulations often rely on GEANT4\cite{GEANT4}. However, muon radiography of large objects requires extensive statistics, making Monte Carlo tools like CORSIKA and CRY too time-consuming for practical use. Muon tomography, being highly sensitive to angular distributions, demands precise simulation of these distributions. Both techniques are typically tailored to specific scenarios, requiring further simulations of muon interactions with objects using GEANT4. For large objects with known structures, simulating muons passing through these structures repeatedly is inefficient. A direct simulation of the muon distribution after traversing known materials could greatly improve computational efficiency.

In response to these challenges, we developed DeepMuon, a data-driven deep learning model for muon generation. DeepMuon is capable of learning various muon distributions, including both simulated and real data, and can rapidly generate muons with the same distribution. Unlike the Monte Carlo method, which relies on sampling from a theoretically computed multidimensional probability density function (PDF), or parameter-based models that use semi-empirical formulas, DeepMuon directly learns from actual cosmic-ray muon data. This approach allows us to more accurately simulate the true distribution of muons, thereby improving the precision of muon tomography, which is particularly sensitive to angular distributions. Additionally, DeepMuon can learn the distribution of muons after they pass through various complex structures, significantly enhancing the simulation speed when using tools like GEANT4.

Deep learning models have extensive applications in high-energy physics\cite{dlexp1,dlexp2,cheng2024application}. Traditional deep learning-based event generation algorithms typically employ GANs (Generative Adversarial Networks)\cite{gan} or VAEs (Variational Autoencoders)\cite{vae}. GANs involve a data generator and a discriminator; the generator creates simulated data, while the discriminator distinguishes between simulated and real data. This adversarial process helps generate highly realistic data. VAEs, on the other hand, encode input data into a latent space and then decode it to generate data, relying on carefully designed loss functions and complex model architectures. In contrast, the Sliced Wasserstein Distance (SWD) offers a more efficient and precise approach to deep learning. SWD measures the distance between two probability distributions by projecting them onto multiple one-dimensional subspaces and calculating the Wasserstein distance (also known as Earth Mover's Distance) in these subspaces. This method captures the differences between distributions with high accuracy and requires less computational effort than directly comparing the distributions in their original high-dimensional space\cite{ot}. Models using SWD as a loss function have been shown to produce high-energy physics events with exceptional precision\cite{pan2023eventgenerationconsistencetest}. DeepMuon leverages a simple transformer encoder paired with the SWD loss function to achieve high-precision fitting of cosmic-ray muon distributions, further validating the powerful distribution-fitting capabilities of SWD.

Cosmic-ray muon energy distributions are sharply peaked, making them challenging for deep learning models to learn. To overcome this, we applied the inverse Box-Cox transformation to smooth the energy distribution, facilitating more effective learning.

In our study, we trained the model using muon data generated by the Cosmic-ray Shower Library (CRY) at sea level. As a demonstration of deep learning's potential to accelerate GEANT4 simulations, we built a pipeline with DeepMuon as the muon source and simulated muon distributions at a depth of 50 meters underwater using GEANT4. We compared these results with simulations conducted using CRY and GEANT4 directly, and our pipeline significantly improved simulation speed while maintaining accuracy, offering a promising solution for underwater and underground muon tomography.

The following sections will detail the architecture of our model, the methods used for preprocessing muon data, and how we leveraged the model to accelerate GEANT4 simulations. We will also present the model's performance and the accuracy of its muon distribution fits.

\section{Method}

\subsection{Data Preprocessing}
To train our cosmic muon generation model, we utilized cosmic muon distribution data generated by CRY to construct our training dataset. Given that the spatial distribution of cosmic muons at sea level is approximately uniform, we focused on using only their energy and angular distributions for model training.

The energy distribution of cosmic muons is highly peaked, as shown in Figure 1, which illustrates the energy spectrum generated by CRY. This sharp distribution creates steep gradients, making it challenging for the model to learn directly from the data. Additionally, the figure reveals a step-like feature in the energy spectrum produced by CRY. This artifact is likely caused by the discretization introduced by CRY’s use of precomputed input tables derived from MCNPX, from which cosmic ray samples are drawn. Similar characteristics have been observed in other studies that use CRY for muon generation.\cite{crydistribute}.
\begin{figure}[H]
    \centering
    \includegraphics[width=0.5\linewidth]{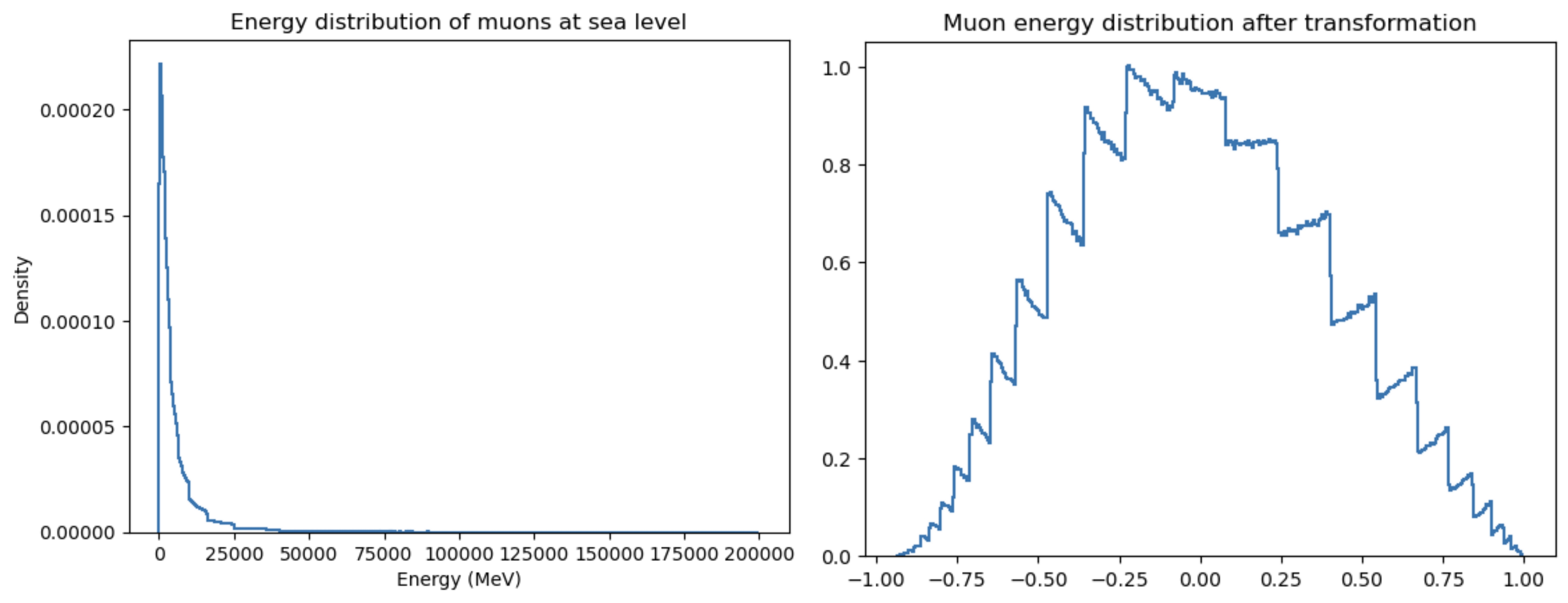}
    \caption{The distribution of cosmic ray muons at sea level generated by CRY is highly skewed (left figure). After applying the inverse Box-Cox transformation, the distribution exhibits improved statistical characteristics (right figure).}
\end{figure}
To address this challenge, we introduced the Box-Cox transformation, a commonly used statistical technique aimed at adjusting the shape of a data distribution. The Box-Cox transformation helps improve the statistical properties of the data, making it closer to a normal distribution. In practice, this transformation is often applied to data with high kurtosis and skewness, which can otherwise lead to misleading results during analysis\cite{box-cox}. The primary goal of the Box-Cox transformation is to modify the shape of the data distribution, making it more suitable for subsequent statistical analysis and modeling. The Box-Cox transformation is defined as follows:

\begin{equation}
    y(\lambda) = \begin{cases} 
        \frac{x^\lambda - 1}{\lambda}, & \text{if } \lambda \neq 0 \\
        \log x, & \text{if } \lambda = 0 
    \end{cases}
\end{equation}
Here, \( x \) represents the input data, and \( \lambda \) is an adjustable parameter. In our approach, we apply the inverse Box-Cox transformation to the energy data. The inverse transformation is defined as follows:

\begin{equation} 
x = 
\begin{cases} 
(\lambda y(\lambda) + 1)^{1/\lambda}, & \text{if } \lambda \neq 0 \\ 
e^{y(\lambda)}, & \text{if } \lambda = 0 
\end{cases} 
\end{equation}

Through experimentation, we found that setting the Box-Cox transformation parameter \( \lambda \) to 9 yields the most effective transformation results. After applying this transformation, the energy data requires further processing to meet the model's input requirements. First, we subtract the mean from the energy values. This step reduces the magnitude of the energy and ensures that its distribution is as symmetric as possible. Symmetric data distributions make it easier for the model to learn the underlying features of the data. Additionally, a symmetric distribution helps mitigate potential training issues such as vanishing or exploding gradients.

After mean subtraction, we apply the hyperbolic tangent (tanh) function to map the energy values into the range (-1,1). This step compresses the input data into a finite range, contributing to the model's stable training. Once the energy data is mapped to the (-1,1) interval, we combine the processed energy values with the directional angles to form a feature vector. This feature vector is then fed into the neural network model for further processing.

\subsection{Model Architecture}
Figure 3 illustrates the overall structure of our muon generation model. We designed a straightforward architecture consisting of a linear embedding layer, a two-layer, eight-head Transformer encoder, and an MLP decoder. The objective of DeepMuon is to transform a uniform distribution into the desired cosmic ray muon distribution. To achieve this, we use a uniform distribution within the range of (-1, 1) as the model's input, with a size of Cx100, where C represents the batch size. This uniform distribution is first transformed by the linear embedding layer into a (Cx3x1024) size, which serves as the input for the Transformer encoder.

The Transformer encoder processes this input and passes it to the MLP decoder, which outputs the multidimensional distribution of cosmic ray muons. The output of the model has dimensions (Cx1024x3), where Cx1024 corresponds to the number of generated cosmic ray muon instances, and the final dimension represents the energy and two zenith angles of each muon instance.

To accurately replicate the muon distribution generated by CRY, we used a dataset of 100 million sea-level cosmic ray muon instances produced by CRY for training. During training, a batch of muon data is randomly sampled from this dataset, and the Sliced Wasserstein Distance (SWD) Loss is calculated to update the model parameters.
\begin{figure}[H]
    \centering
    \includegraphics[width=0.8\linewidth]{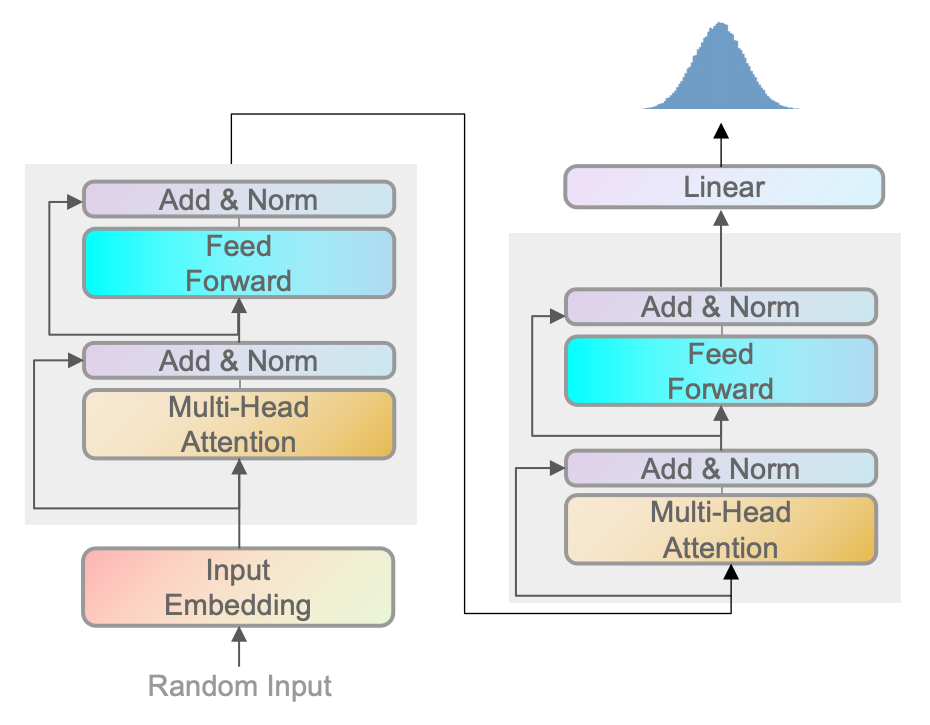}
    \caption{The structure of the DeepMuon model: A uniform distribution is provided as input, which is projected by a linear embedding layer to match the input dimensions of the Transformer encoder. The Transformer encoder then encodes this into a high-dimensional latent space, and finally, the linear decoder transforms it into the distribution of cosmic ray muons.}
\end{figure}
\subsection{Model Performance}
Figures 3-4 present the performance of our sea-level muon generation model. We compared the energy distributions of muons generated by our model with those produced by CRY, and also examined muon fluxes at different zenith angles (0° and 60°). The distribution plots show a high degree of consistency between the data generated by our model and the CRY-generated target data.

It is important to note that deep learning models, by design, are continuous and differentiable functions, whereas CRY's output exhibits a step-like pattern, which does not accurately reflect the true distribution of cosmic muons. Although our model can, over the course of training, learn to replicate these step-like features, we consider this an overfitting to the undesirable characteristics of the CRY data. To avoid this, we halt training before the model overfits to CRY's step-like artifacts.
\begin{figure}[H]
    \centering
    \includegraphics[width=0.8\linewidth]{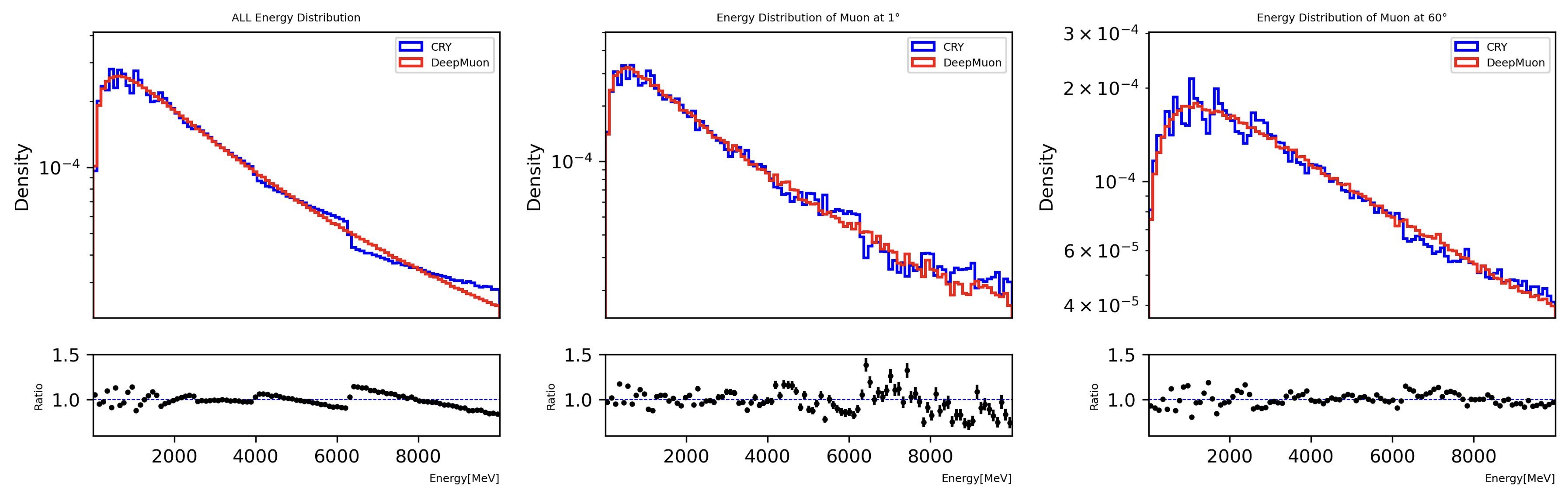}
    \caption{From left to right show the energy distribution for all events, the energy distribution of muons at a 1° zenith angle, and the energy distribution of muons at a 60° zenith angle, respectively.}
\end{figure}
\begin{figure}[H]
    \centering
    \includegraphics[width=0.8\linewidth]{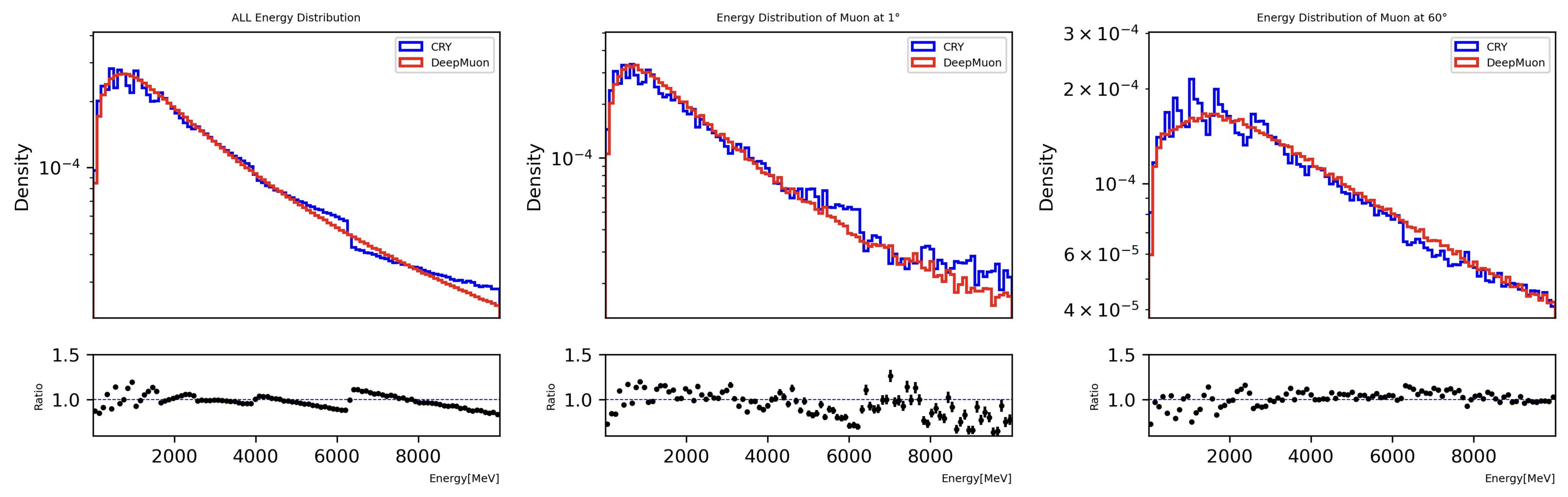}
    \caption{Even with a limited amount of data, DeepMuon is capable of effectively learning the distribution of cosmic ray muons.}
\end{figure}

In contrast to simulated data, real-world data is often more limited. We do not have access to large datasets of real cosmic muons comparable to the 100 million samples generated by CRY for training. To address this, we tested our model using a much smaller subset of CRY-generated data (6,000 samples) and compared the model's output with CRY's full dataset (100 million samples). As shown in the figures, even with a limited amount of training data, our model was able to accurately capture the correct distribution. This suggests that our model is capable of learning the true cosmic muon distribution even when working with smaller, real-world datasets.

We evaluated the inference speed of our model using an NVIDIA RTX 3090 GPU. In our tests, DeepMuon generated 20.48 million cosmic ray muons in just 16 seconds. By comparison, generating the same number of sea-level cosmic ray muons with CRY on a CPU required 4,880 seconds. This demonstrates that DeepMuon achieves a speedup of up to 288 times over the CPU. This significant acceleration is particularly beneficial for applications like muon absorption radiography, where rapid generation of large datasets is crucial.
\subsection{Simulation acceleration}

In scenarios involving the imaging of large objects, fully simulating the passage of cosmic ray muons through such objects is often time-consuming. The purpose of imaging large structures is typically to investigate specific internal features or material properties. Our primary focus is on how these features affect the distribution of emerging cosmic ray muons. Therefore, when certain fixed structures inside the object are already known, repeatedly simulating them is unnecessary. By leveraging our deep learning model, we can simulate cosmic ray muon fluxes with arbitrary distributions. This allows us to directly learn and generate muon distributions that have passed through the fixed structures, significantly accelerating the simulation process for large-object imaging.
\begin{figure}[H]
    \centering
    \includegraphics[width=0.6\linewidth]{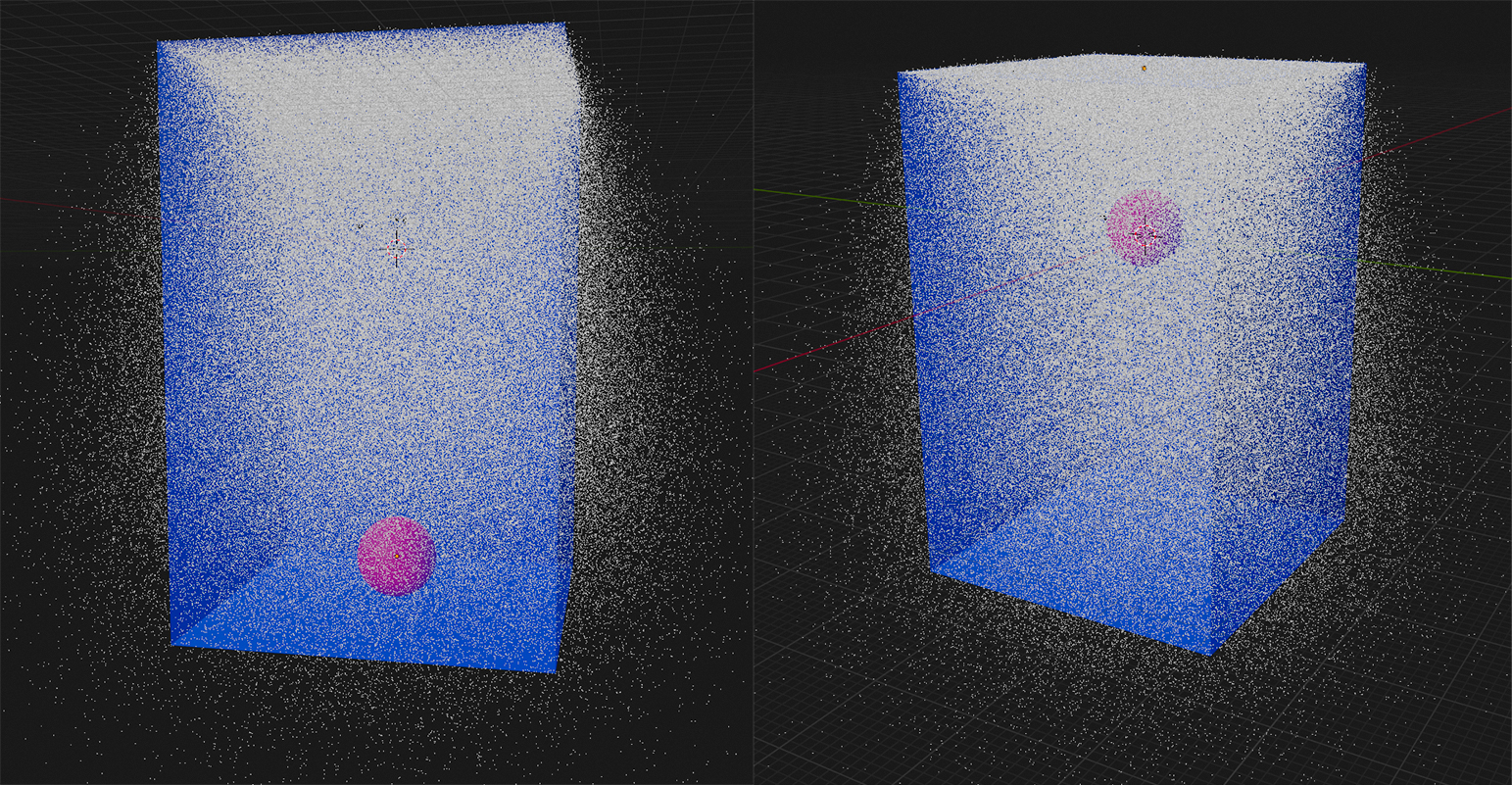}
    \caption{Simulating muons at greater depths in GEANT4 significantly reduces the flux and affects the angular distribution.}
\end{figure}
As illustrated in Figure 5, consider the example of imaging an underwater sphere using muon scattering tomography. To achieve this, we would need to simulate the entire process of muons passing through both the water and the sphere. However, in practice, our primary concern is the sphere's effect on the muon distribution. Simulating the water layer above the sphere is not only time-consuming but also reduces the muon flux reaching the sphere. If we can directly generate the muon distribution beneath the water, we can effectively transform the deep-water imaging problem into a shallow-water one, drastically shortening simulation time. In this chapter, we demonstrate how we use the DeepMuon model to build a pipeline that accelerates this simulation process by generating the underwater muon distribution.

To simulate the underwater muon distribution, we created a cubic water-filled world in GEANT4. We emitted cosmic ray muons from the top and simulated their passage through the water. However, as the depth increases, the number of simulated events in GEANT4 rises significantly. Additionally, to obtain a realistic angular distribution at greater depths, the area of the muon source at the surface must be expanded. For example, to simulate muon distribution at a depth of 50 meters, using an area of 90,000 m² (corresponding to a GEANT4 world size of 300m x 300m x 50m) results in a maximum zenith angle of about 83°. In contrast, simulating the distribution at only 10 meters depth requires a much smaller world size of 60m x 60m x 10m to achieve the same maximum zenith angle.

Since the rate of muon generation is constant, increasing the surface area drastically reduces the muon flux. Even when accounting for the acceleration of muon generation on GPUs—up to 7,200 times faster than a single CPU core—the problem persists. As depth increases, some muons exit the simulation boundary, further reducing flux and affecting angular distribution, thus compromising simulation accuracy. Therefore, instead of using CRY to generate surface muons and simulating their complete passage through water in GEANT4, we developed a pipeline using DeepMuon to accelerate the simulation of underwater muon distribution. DeepMuon directly generates muon distributions at any given depth, while GEANT4 only simulates the remaining, shallower layers. This approach not only reduces the number of events to be computed but also mitigates the issue of flux reduction with increasing depth.

Specifically, we first train our model using surface muon data generated by CRY, simulating their passage through water in GEANT4 to obtain underwater muon distributions. We then use this data to further train a model for underwater muon generation, repeating this process to create models for different depths. The pipeline consists of two key steps:
\begin{enumerate}
    \item Training DeepMuon to learn the underwater muon distribution simulated by GEANT4.
    \item Integrating DeepMuon into GEANT4 to complete the simulation for deeper underwater muon distributions.
\end{enumerate}
Specifically, we first trained the DeepMuon model using muon distribution data generated by CRY at sea level. Once the DeepMuon muon generator was established, we used it as input for the particle gun in GEANT4. In GEANT4, we created a 60m x 60m x 10m simulation space filled with water. The particle gun randomly emitted muons from the top of this volume, while a muon detector at the bottom recorded the distribution of muons that had passed through 10 meters of water. After obtaining this distribution data, we used it to train a new DeepMuon model for deeper water conditions, and repeated the process by using this updated model as input for the next simulation step in GEANT4.

By progressing through these layers step by step, we effectively transformed the problem of deep-water cosmic muon simulation into multiple faster, shallow-water simulations. Leveraging our model's ability to learn arbitrary cosmic muon distributions, we significantly increased the simulated muon flux while avoiding unnecessary computations, thereby greatly improving the efficiency of cosmic muon simulation.
\subsubsection{Result}
Using the proposed pipeline, we developed DeepMuon models capable of generating muon distributions at varying underwater depths: 0m, 10m, 20m, 30m, 40m, and 50m. For comparison, we used the CRY tool to generate sea-level muons and then simulated their propagation through 10 meters and 50 meters of water using GEANT4. We then compared these results with the distributions produced by our 10m and 50m DeepMuon models. The comparison results are shown in Figures 6-7.
\begin{figure}[H]
    \centering
    \includegraphics[width=0.8\linewidth]{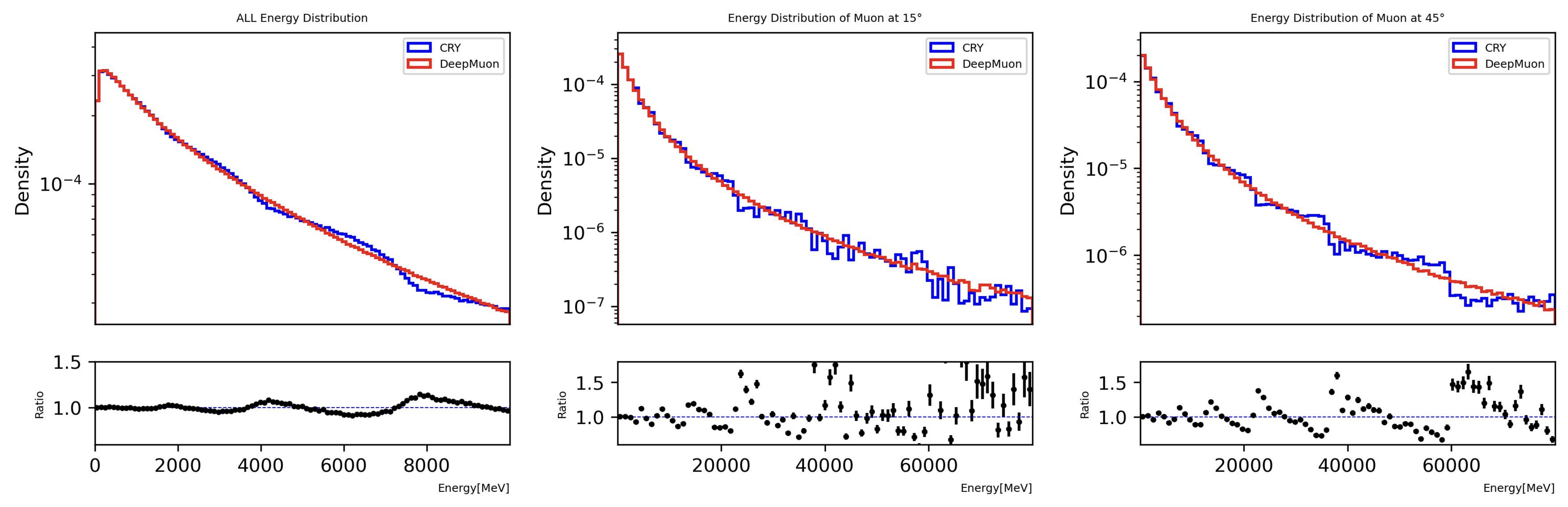}
    \caption{Energy distribution of cosmic ray muons at a depth of 10 meters underwater.}
    \label{fig:enter-label}
\end{figure}

\begin{figure}[H]
    \centering
    \includegraphics[width=0.8\linewidth]{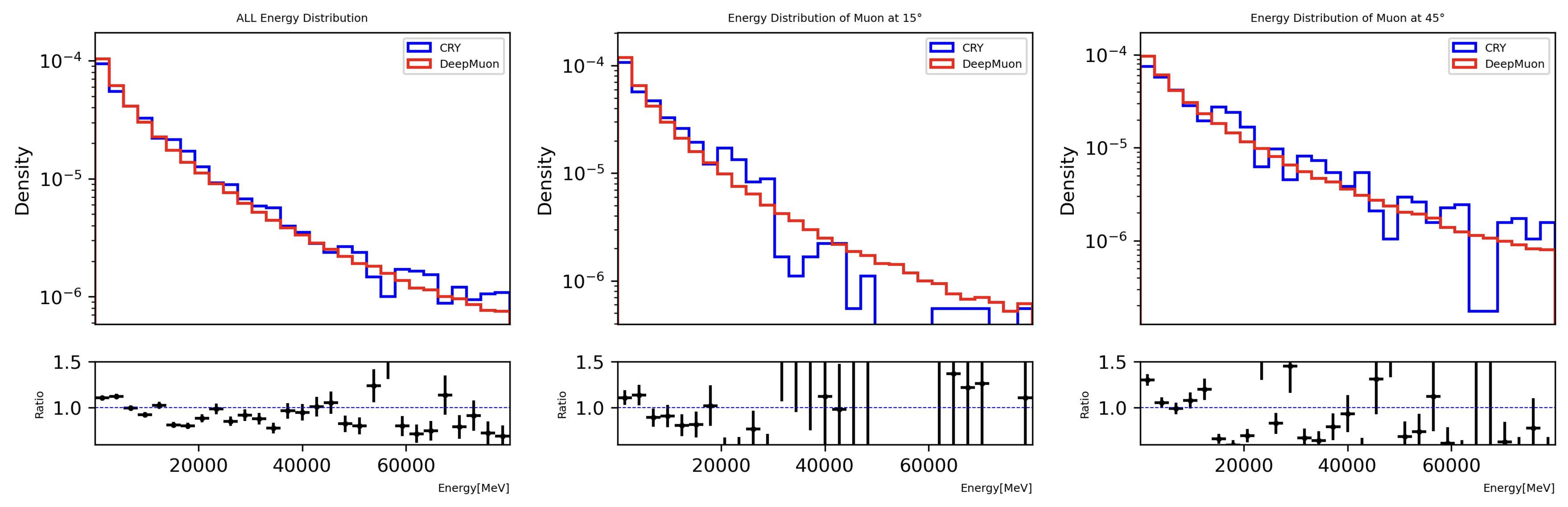}
    \caption{Energy distribution of cosmic ray muons at a depth of 50 meters underwater.}
\end{figure}

At a depth of 10m, the CRY-generated muon data maintained sufficient statistical significance, and the distribution produced by the DeepMuon model closely matched the CRY+GEANT4 simulation. However, at 50m depth, the muon flux reaching the detector dropped significantly, leading to extended simulation times and reduced data output from CRY+GEANT4, making it difficult to observe a smooth, complete distribution. Although only a limited number of events could be generated for comparison, the DeepMuon model still demonstrated a strong agreement with the distribution trend, indicating that the model’s output remained consistent and did not introduce significant distortions during the pipeline's propagation.

\section{Conclusion}
DeepMuon is an efficient method for the rapid simulation of cosmic ray muon distributions. By applying the inverse Box-Cox transformation, DeepMuon reduces the kurtosis of cosmic ray muon energy distributions, bringing them closer to a normal distribution. This transformation makes the data more accessible for deep learning models to process. Additionally, the use of the Sliced Wasserstein Distance (SWD) loss function allows the model to directly learn from the distribution data of cosmic ray muons without the need for added physical constraints to achieve accurate simulations.Our results demonstrate that DeepMuon achieves high precision in simulating cosmic ray muon distributions across various scenarios, all while maintaining remarkable computational speed. This means DeepMuon can replace traditional Monte Carlo muon generators, offering faster and more accurate cosmic ray muon distribution simulations. Moreover, by learning the distribution of muons passing through specific structures, DeepMuon accelerates processes in muon radiography and tomography. The model significantly reduces CPU time, enabling simulations that were previously challenging, such as those involving deep-sea or deep-well muon distributions.We also show that DeepMuon can effectively learn muon distributions from a limited amount of training data. This indicates the model's potential to directly learn from real data collected by muon detectors, which could lead to even higher simulation accuracy. As a distribution transformation model based on optimal transport loss functions, DeepMuon demonstrates the flexibility to convert uniform distributions into various cosmic ray muon distributions. This suggests broader applications in future work, where DeepMuon could not only serve as a generation model but also be used to model the effects of structures like water layers or rock formations on muon distributions. Furthermore, it could be applied to study the impact of particle detectors on high-energy particle beams in physics experiments.In conclusion, DeepMuon offers a transformative approach to cosmic ray muon simulation, combining high precision, speed, and versatility, opening new possibilities for both applied and theoretical research in muon radiography, tomography, and particle physics.
\section{Declaration of generative AI and AI-assisted technologies in the writing process}
During the preparation of this work the authors used ChatGPT in order to translate. After using ChatGPT, the authors reviewed and edited the content as needed and take full responsibility for the content of the publication.

\bibliographystyle{elsarticle-num} 
\bibliography{ref}






\end{document}